\documentclass[10pt,twocolumn,letterpaper]{article}
\usepackage[pagenumbers]{cvpr} 

\usepackage[dvipsnames]{xcolor}

\usepackage{graphicx}
\usepackage{amsmath}
\usepackage{amssymb}
\usepackage{booktabs}
\usepackage[accsupp]{axessibility}  
\definecolor{cvprblue}{rgb}{0.21,0.49,0.74}
\usepackage[pagebackref,breaklinks,colorlinks,citecolor=cvprblue]{hyperref}

\def\paperID{3028} 
\def\confName{CVPR}
\def\confYear{2024}

\newcommand{\ft}[1]{\textcolor{black}{#1}}

\title{DreamHead: Learning Spatial-Temporal Correspondence via Hierarchical Diffusion for Audio-driven Talking Head Synthesis}

\author{Fa-Ting Hong$^{1,2}$ \and Yunfei Liu$^2$ \and Yu Li$^2$ \and Changyin Zhou$^3$ \and Fei Yu$^3$ \and Dan Xu$^1$
\vspace{3pt}
\and
$^1$Department of Computer Science and Engineering, HKUST\\
\vspace{1pt}
$^2$International Digital Economy Academy \quad  $^3$Vistring \\
{\tt\small fhongac@connect.ust.hk, liuyunfei@idea.edu.cn, liyu@idea.cn, } \\
{\tt\small chanyin.zhou@gmail.com, yufei.flyingfish.gmail.com, danxu@cse.ust.hk}
}
\begin{document}

\twocolumn[{%
\renewcommand\twocolumn[1][]{#1}%
\maketitle
\begin{center}
    \centering
    \captionsetup{type=figure}
    \vspace{-8pt}
    \includegraphics[width=0.99\textwidth]{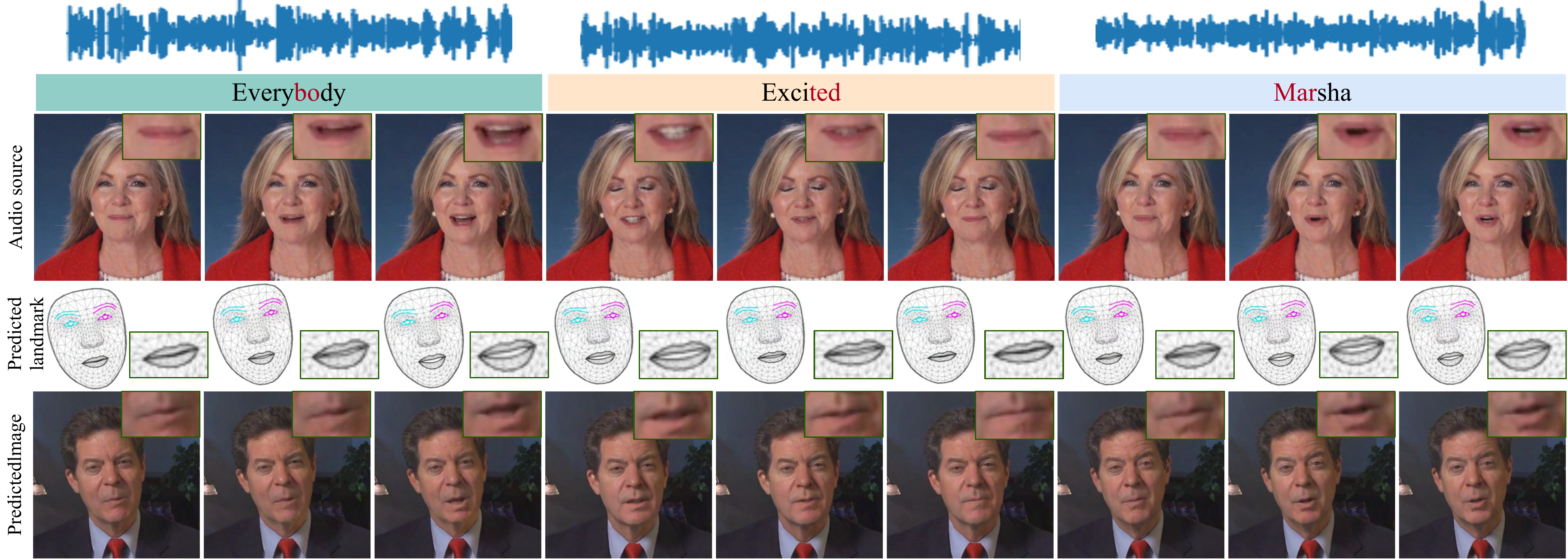}
    \vspace{-5pt}
    \captionof{figure}{We propose a hierarchical diffusion framework (DreamHead) that learns to diffuse facial landmarks as intermediate signals to represent facial expression and targets learning spatial-temporal correspondences in audio-driven talking head synthesis. Given a driving audio sequence, our DreamHead can estimate jittering-less landmark sequences corresponding to the audio temporally and synthesize temporal-smooth and lip-synced talking videos via explicit spatial consistency from predicted landmarks, and no GT landmarks are required during inference. 
    }
    \label{fig:firstpage}
\end{center}%
}]

\begin{abstract}
\vspace{-10pt}
Audio-driven talking head synthesis strives to generate lifelike video portraits from provided audio. The diffusion model, recognized for its superior quality and robust generalization, has been explored for this task. However, establishing a robust correspondence between temporal audio cues and corresponding spatial facial expressions with diffusion models remains a significant challenge in talking head generation.
To bridge this gap, we present DreamHead, a hierarchical diffusion framework that learns spatial-temporal correspondences in talking head synthesis without compromising the model's intrinsic quality and adaptability.~DreamHead learns to predict dense facial landmarks from audios as intermediate signals to model the spatial and temporal correspondences.~Specifically, a first hierarchy of audio-to-landmark diffusion is first designed to predict temporally smooth and accurate landmark sequences given audio sequence signals. Then, a second hierarchy of landmark-to-image diffusion is further proposed to produce spatially consistent facial portrait videos, by modeling spatial correspondences between the dense facial landmark and appearance. Extensive experiments show that proposed DreamHead can effectively learn spatial-temporal consistency with the designed hierarchical diffusion and produce high-fidelity audio-driven talking head videos for multiple identities. 
\end{abstract}
    
\section{Introduction}
\vspace{-3pt}

Talking head generation~\cite{chung2017lip,siarohin2019first,hong2022depth,hong2023implicit,liu2023moda, hong2023dagan++, zhou2021pose} aims to produce realistic-looking portrait videos with a specific driven condition. Given an audio-driven signal, talking head generation methods synthesize facial images with lip shapes synchronized to the audio. Audio-driven talking head methods can be widely applied in various applications, including live broadcasting and video conferencing. 

\par {In recent developments, generative adversarial networks (GANs) ~\cite{goodfellow2020generative, prajwal2020lip, zhou2021pose, liu2023moda} produce portrait videos with limited quality because of their inherent instability in the training stage. Meanwhile, diffusion models~\cite{rombach2022high, shen2023difftalk, stypulkowski2023diffused,du2023dae} become popular because they can render high-quality images while maintaining their generalization capabilities. However, the performance of existing diffusion models~\cite{shen2023difftalk, stypulkowski2023diffused,du2023dae} for taking head generation falls short of expectations. \ft{Using diffusion model to address talking head video generation still remains challenging and is non-trivial to realize high-quality generation.}~A remaining major challenge is that the existing diffusion-based models fail to explicitly establish accurate spatial-temporal correspondences between the audio input and the target facial dynamics. Instead, current methods~\cite{shen2023difftalk} predominantly rely on the diffusion model's implicit learning capabilities, which proves insufficient for effectively capturing and synchronizing the audio cues with the facial expressions. To this end, we propose to model the spatial-temporal correspondence between the given audio sequence and the target facial image sequence within a diffusion model.  \ft{Because facial dynamics lie on a high-dimensional manifold, making it nontrivial to find a mapping from audio~\cite{zhou2020MakeItTalk}. }\ft{However, we found that landmarks serve as an efficient representation for facial expressions \cite{zhou2020MakeItTalk,lu2021LSP,liu2023moda}. Therefore, we employ two diffusion stages to map the audio to lip animation, utilizing landmarks as an intermediate representation to bridge two stages. Taking the landmarks as intermediate representation, the two-stage strategy allows us to alleviate the challenges associated with learning audio-lip mapping}.}

\par

To tackle the above-discussed problem for high-fidelity talking head video generation, specifically,
we present a novel hierarchical diffusion framework, termed as \textbf{DreamHead}.~It is designed to learn the spatial-temporal correspondence between the audio sequence and facial dynamics for synthesizing portrait videos. Inspired by ~\cite{zhou2020MakeItTalk,lu2021LSP,liu2023moda}, we realize that dense facial landmarks can represent facial expressions accurately.
{To model the temporal information contained in the audio input,} in the first hierarchy of diffusion, we design a lightweight audio-to-landmark diffusion (A2L) for learning the temporal correspondence between the given audio and the facial landmarks sequence.~With a diffusion-based optimization,
our designed audio-to-landmark model can eliminate the jittering artifacts and estimate temporal-smooth landmark sequences.~In the second hierarchy of diffusion, 
{
we further design a landmark-to-image diffusion (L2I) to generate photo-realistic portrait videos with the facial landmarks predicted from the first hierarchy of diffusion (\emph{i.e.}, A2L). 
With the predicted dense facial landmarks and their corresponding face images, the designed L2I diffusion can learn to effectively model spatial correspondence between the landmarks and the face expression with self-attention aggregation in the diffusion process. Therefore, with the proposed hierarchical diffusion model, we can effectively construct spatial and temporal correspondences between the input audio signal and the output face images through the intermediate facial landmarks, leading to a higher-quality generation of cross-modal synchronized faces. Moreover, as the facial landmarks are generated through the first hierarchy of diffusion (\emph{i.e.}, A2L), no ground-truth facial landmarks are required during the inference stage.  
}

\par We conduct extensive experiments to evaluate the proposed DreamHead on two competitive audio-driven talking head generation datasets (\ie HDTF~\cite{zhang2021HDTF} and MEAD~\cite{wang2020mead}). From the experimental results, we can observe that our designed audio-to-landmark diffusion can estimate accurate and stable facial landmark sequences, given the audio condition. The landmark-to-image diffusion can generate realistic appearance and spatial-temporal consistent portrait video. The experimental results also show clearly improved generation results over state-of-the-art methods from both qualitative and quantitative perspectives.

In summary, our main contribution is three-fold:
\begin{itemize}
    \item {We propose learning dense facial landmarks as an intermediate bridge within a diffusion framework to model the spatial-temporal correspondence for talking head synthesis. Integrating landmarks can ensure temporal consistency between the audio and the final video, and provides explicit spatial constraints for accurate lip movements in synthesized talking head videos.}
    \item We design a novel hierarchical diffusion model, coined as DreamHead, for audio-driven talking head video generation.~Specifically, DreamHead is constructed by a hierarchy of two diffusion structures:~(i) Audio-to-landmark diffusion structure generates accurate and temporal-smooth jittering-less facial landmarks from the input audio.~(ii) Landmark-to-image diffusion structure synthesizes high-fidelity portrait videos via learning spatial correspondences between landmarks and facial expressions.

    \item Extensive experiments on two datasets show that DreamHead can estimate jitter-less landmark sequences, and produce spatially and temporally consistent portrait videos. Furthermore, our diffusion model exhibits superior generation performance across the different benchmarks compared to state-of-the-art counterparts.
\end{itemize}

\vspace{-3pt}

\section{Related Works}
\vspace{-3pt}
\label{sec:related}
\noindent\textbf{Audio-driven Talking Head Generation.} 
Existing approaches \cite{garrido2015vdub,hong2024learning,chung2017lip,suwajanakorn2017synthesizing,taylor2017deep,zhou2019talking} can be divided into two categories, \ie subject-specific and subject-general.~Regarding subject-specific audio-driven talking head methods~\cite{guo2021adnerf,liu2023moda, ye2023geneface,lu2021LSP,ye2023geneface++, du2023dae}, they train a personalized renderer for each person. For instance, MODA~\cite{liu2023moda} learns the correlation between lip shape and other movements on the face to synthesise a natural face. AD-NeRF~\cite{guo2021adnerf} attempts to leverage volume rendering on two elaborately designed NeRFs to synthesize portrait videos directly.
Different from those subject-specific methods, subject-general methods~\cite{ji2022eamm,gururani2023space,sun2022masked, zhou2021pose,prajwal2020lip,zhou2020MakeItTalk,tang2022memories,zhou2019talking,shen2023difftalk} can produce portrait videos given audios and human faces with multiple identities. Wav2Lip~\cite{prajwal2020lip} accepts an audio sequence as input and inpaint the mouth region while maintaining other facial regions unmodified. Different from Wav2Lip~\cite{prajwal2020lip}, PC-AVS~\cite{zhou2021pose} tried to disentangle the pose and identity representations from a face image to enable the control of face generation. Besides,SadTalker~\cite{zhang2022sadtalker} uses audio to estimate 3D coefficients for talking head generation. Different from the above-discussed methods, our proposed DreamHead framework focuses on subject-generic audio-driven talking head generation. It leverages the diffusion process to produce high-quality face video generation and achieve the capability of model generalization. 

\noindent\textbf{Latent Diffusion Models.} 
{
Latent Diffusion Models~\cite{rombach2022high} has gained attention in recent years because of their strong ability in various image generation tasks~\cite{lugmayr2022repaint,qi2023fatezero,shen2023difftalk,du2023dae,ho2022imagen,karras2023dreampose,inoue2023layoutdm}, \eg, video editing~\cite{qi2023fatezero} and motion transfer~\cite{karras2023dreampose}. Because of the stable training procedure and high-quality generation, researchers also utilize diffusion models to deal with the talking head generation task~\cite{shen2023difftalk,stypulkowski2023diffused}. For instance, DiffTalk~\cite{shen2023difftalk} follows the setting of Wav2Lip~\cite{prajwal2020lip} and designs an autoregressive method to produce portrait video in their inference stage.~DAE-Talker~\cite{du2023dae} utilized a diffusion autoencoder to address subject-specific talking head generation.}~{However, these models rely on implicit learning capabilities of the diffusion model to map the audio to images, which cannot effectively model important spatial relationships for image synthesis. In this work, we design a hierarchical diffusion framework to learn the spatial-temporal correspondence between the input audio and the output face video using diffused landmarks as intermediate guidance signals. In this way, our method can produce spatially consistent and temporally smooth portrait videos.}

\begin{figure*}[t]
  \centering
    \includegraphics[width=1\textwidth]{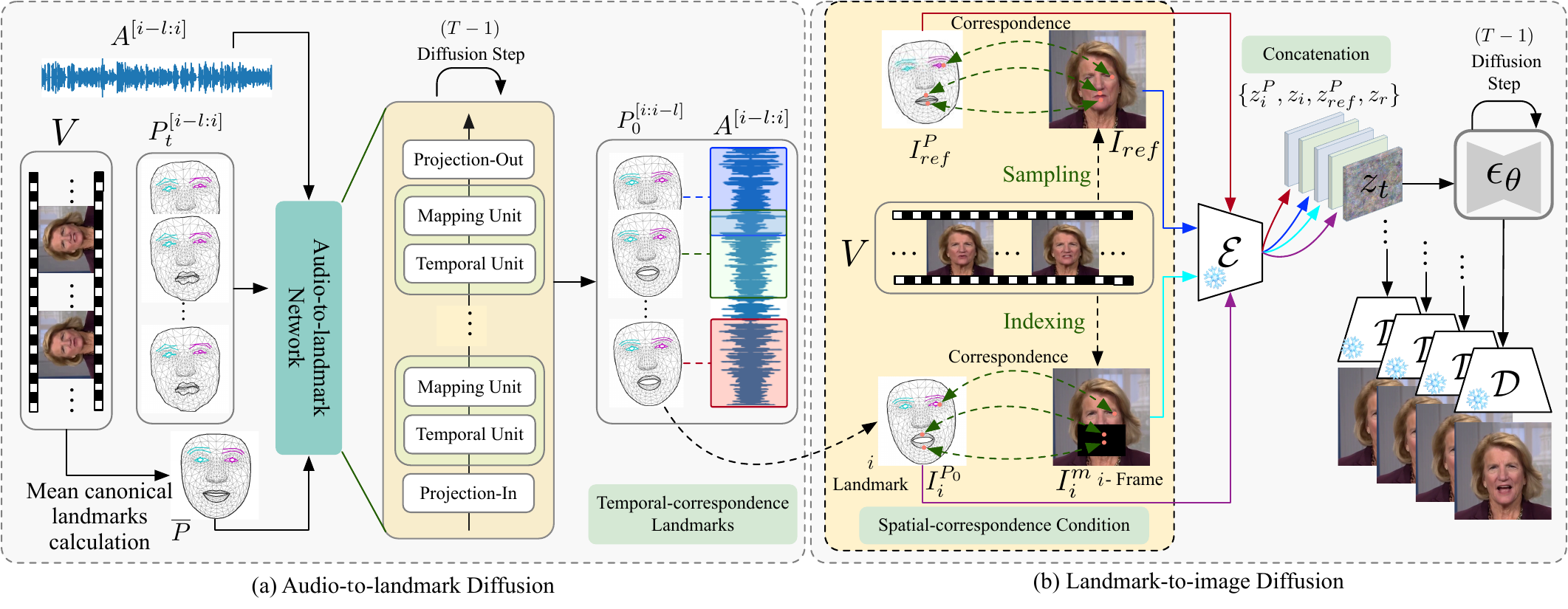}
    \vspace{-20pt}
    \caption{The illustration of our framework. Our proposed DreamHead diffuses landmarks as intermediate signals to learn the spatial-temporal correspondence for talking head video generation. (a) The first hierarchy of audio-to-landmark diffusion (A2L) takes an audio sequence as input to predict a temporal-correspondence landmark sequence with a corresponding lip shape. By cooperating with temporal units in the A2L network, we can produce a jittering-less landmark sequence. (b) The second hierarchy of landmark-to-image diffusion process aims to produce the final portrait video given a spatial-correspondence condition set. $\mathcal{E}$ is an image encoder to downsample the input image, while the $\mathcal{D}$ is a decoder that upsamples a latent to generate an image.
    }
    \vspace{-12pt}
    
    \label{fig:framework}       
\end{figure*}
\vspace{-3pt}

\section{The Proposed DreamHead Approach}
\vspace{-3pt}

\ft{Given a audio sequence $A$ and a video clip $V$, a talking portrait method aims to map it into the corresponding video clip $\tilde{V}$, which should have the length as same as the audio sequence $A$.} Following the previous works~\cite{prajwal2020lip,shen2023difftalk}, we aim to inpaint the lip region corresponding to the given audio while maintaining other parts of the faces unchanged.


\vspace{-3pt}
\subsection{Overview}
\vspace{-3pt}

As illustrated in Figure~\ref{fig:framework}, our DreamHead is composed of a hierarchy of two diffusion processes, which are connected by diffused dense facial landmarks: (i) Audio-to-landmark diffusion stage. Taking the audio context into consideration, DreamHead learns an audio-to-landmark diffusion in the first hierarchy to estimate a facial landmark sequence corresponding to the given temporal audio sequence. Given a video as input, we first compute the mean canonical landmarks $\overline{P}$ and feed it with the given audio sequence into the audio-to-landmark diffusion (\textbf{A2L}) process to \ft{generate the lip-synced landmarks sequences}. 
{Importantly, the A2L diffusion process efficiently aligns audios with landmarks using multiple temporal blocks, ensuring temporally smooth and synchronized landmark sequences.} (ii) In the second hierarchy of landmark-to-image diffusion, we model spatial correspondence between landmarks and face expressions, which allows the landmark-to-image diffusion (\textbf{L2I}) to produce spatially consistent frames with accurate appearance.
Additionally, we draw the landmarks in canvas as images ($I_i^{P_0}$ and $I_{ref}^P$) and feed them into the L2I diffusion model. In this way, the model can learn the spatial correspondence between expression and landmarks. As the generated landmarks through A2L are temporally smooth, the generated final faces can achieve spatial-temporal consistency. In the following, we will show the detailed learning procedures of these two diffusion processes.

\subsection{Audio to Landmark Diffusion}
\vspace{-3pt}

\label{sec:a2l}
Mapping audio to the distribution of pixels is a highly ill-posed problem since it involves dealing with a vast, continuous output space with a high degree of variability and less direct correlation to the input audio signals. Instead, audio-to-landmark mapping is a more constrained problem because of the inherent correspondence between the phoneme and landmarks of the lip.
Therefore, we first learn a diffusion to map the input audio sequence to a temporally smooth landmark sequence using the first hierarchy of audio-to-landmark diffusion of our DreamHead framework. 

\begin{figure*}[t]
  \centering
    \includegraphics[width=1\linewidth]{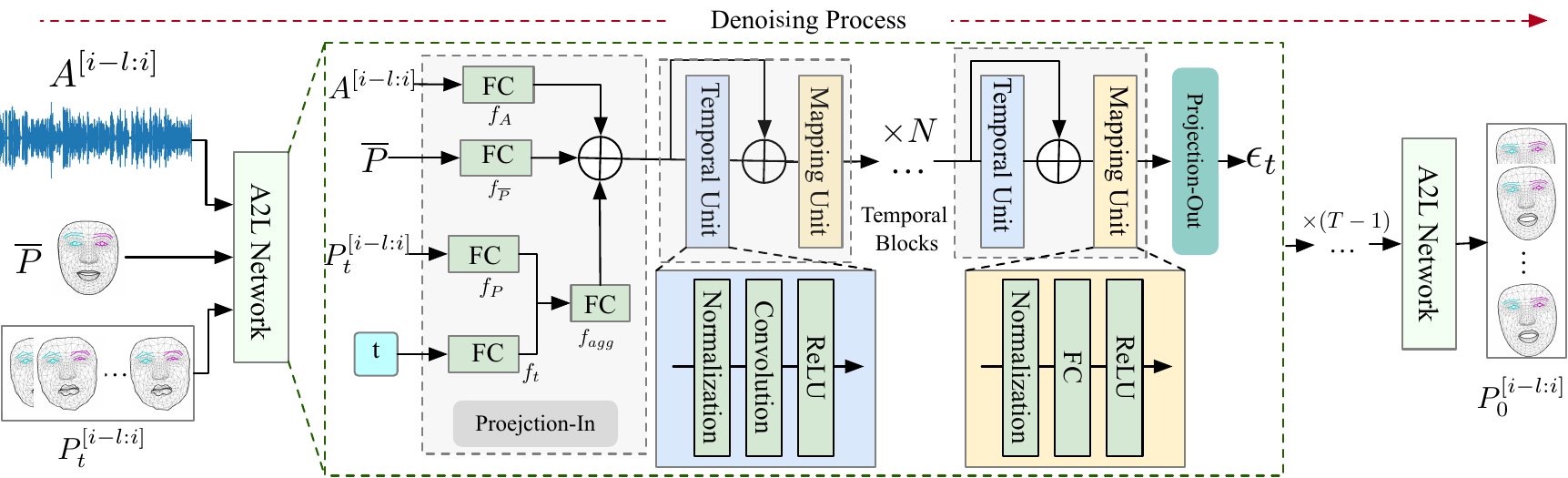}
    \vspace{-20pt}
    \caption{The illustration of our audio-to-landmark diffusion process. The audio-to-landmark network (A2L network) contains multiple fully connected layers to change the dimensions of inputs. Moreover, multiple temporal units in the A2L network can perceive the temporal information from audio cues. 
    }
    \vspace{-10pt}
    
    \label{fig:audio2landmark}  
\end{figure*}

\noindent\textbf{Landmark normalization.} To facilitate the learning of the audio-to-landmark diffusion process, we preprocess the facial landmarks to enhance the connections between the audio cues and the facial landmarks by filtering the audio-irrelevant information in the dense facial landmarks. 
To make the model focus on lip dynamics, we first project the 3D dense facial landmarks into a canonical space to eliminate the influence of head poses. Then, we normalize each canonical landmark by the mean and variance, which are calculated from all the canonical landmarks in the given video. Therefore, even a randomly initialized noise can obtain a rough face outline after de-normalizing (see $P_t^{[i-l:i]}$ in Figure~\ref{fig:audio2landmark}). \ft{Generally, the variance in canonical landmarks can represent the talking style of a given person. Because audio is weakly related to personalized talking styles, we can eliminate the talking style information in the landmarks through normalization to facilitates the learning of the audio-to-landmark mapping. In this way, we train our audio-to-landmark diffusion with normalized pose-irrelevant canonical landmark as supervision.} 


\noindent\textbf{Temporal modeling.}
{To capture the temporal dynamics within the audio sequence, we feed an audio segment that corresponds to image frames from the $(i-l)$-th to the $i$-th in the video into the A2L diffusion.} Before feeding audios into the A2L model, we first process the input audio sequence using the pre-trained wav2vec model~\cite{2019wav2vec}, resulting in a sequence of audio embeddings $A^{[i-l:i]}$.  \ft{\ft{To provide the identity information for landmarks generation}, we also feed the mean facial landmarks $\overline{P}$ of all canonical landmarks in a video as a condition into the first hierarchy of diffusion (see Figure~\ref{fig:audio2landmark}).}
 As shown in Figure~\ref{fig:audio2landmark}, we further feed the audio embeddings through a fully connected layer $f_A(\cdot)$. This adaptation tailors the audio features specifically for the task of talking head generation. Additionally, we also utilize several fully connected layers ($f_{\overline{P}}(\cdot)$, $f_{P}(\cdot)$, $f_t(\cdot)$ and $f_{agg}(\cdot)$) to change the dimension of the mean facial landmark $\overline{P}$, landmark sequence $P^{[i-l:i]}_t$ and the timesteps $t$ as follows:
 \begin{equation}
     X_{a2l} = f_A(A^{[i-l:i]})+f_{\overline{P}}(\overline{P}) + f_{agg}(f_{P}(P^{[i-l:i]}_t) \oplus f_t(t)),
 \end{equation}
where $\oplus$ is the concatenation operator and $f_{agg}$ is used to fuse the features of the timesteps and the landmarks.
To establish the temporal correspondence between the audio and the facial landmarks, we utilize a sequence of temporal blocks, which contains a temporal unit and a mapping unit, inside the A2L diffusion model to explore the temporal correlation of all inputs. As shown in the Figure~\ref{fig:audio2landmark}, we adopt $N$ temporal blocks in A2L network to encode the condition sequentially as follows:
\begin{equation}
    \epsilon_t =f_{proj}^{out} \circ f_{tb}^N \circ f_{tb}^{N-1} \circ \dots \circ  f_{tb}^1(X_{a2l}),
\end{equation}
where $f_{tb}^n$ is the $n$-th temporal block and $f_{proj}^{out}$ is the projection-out function as shown in Figure~\ref{fig:audio2landmark}, which is a fully-connected layer. \ft{The ``$\circ$'' denotes module composition, indicating the output of the latter is the input of the former}. To connect the temporal context of the input audio embeddings with a constraint on parameter growth, we implement the temporal unit as a combination of a normalization layer, a convolutional layer \cite{bar2024lumiere}, and a ReLU function, while the mapping unit is a combination of a normalization layer, a fully connected layer, and a ReLU function to smooth the output (see Figure~\ref{fig:audio2landmark}). By stacking multiple temporal blocks, our audio-to-landmark network aggregate the temporal cues from the audio and produce temporal-smooth and jitter-less facial landmark sequences with limited parameters.

\noindent\textbf{Diffusion learning.} Instead of directly predicting the landmarks given an audio sequence, we employ a diffusion process to align the landmark sequence with the audio sequence gradually. We build our audio-to-landmark network as a time-conditional denoising network, which learns the reversion process of a Markov Chain~\cite{Geyer_1992} of length $T$. The corresponding objective can be formulated as follows:
\begin{equation}
    L_{a2l} := \mathbb{E}_{P_t, \epsilon\sim\mathcal{N}(0,1),t}\left[||\epsilon-A2L(P_t^{[i-l:i]},t, C_{a2l})||\right],
\end{equation}
{where A2L$(P_t^{[i-l:i]},t, C_{a2l})$ represents the noise prediction produced by the audio-to-landmark diffusion,}
and $C_{a2l}$ is the condition set \{$A^{[i-l:i]}$, $\overline{P}$\}.
By modeling the audio-to-landmark mapping as a cross-modal diffusion process, we can align the audio and the landmarks well to produce an accurate and temporally smooth facial landmark sequence with less jittering artifacts.

\vspace{-3pt}
\subsection{Landmarks to Image Diffusion}
\vspace{-3pt}

The facial landmarks contain accurate lip shapes estimated by the first hierarchy of the audio-to-landmark (A2L) diffusion. These landmarks provide explicit spatial structure cues about face expressions, which are very beneficial for producing lip-synced face videos. To ensure a high-quality generation of the face videos and achieve a good generalization ability of the model, we adopt a landmark-to-image (L2I) diffusion process, which is a latent diffusion model~\cite{rombach2022high}, to generate the portrait videos in the second hierarchy of our landmark-to-image diffusion. 

\noindent\textbf{Explicit spatial cues from facial landmarks.}~As shown in Figure~\ref{fig:framework}(b), given a target masked image $I_i^{m}$ that provides the appearance, head pose, and image background information, our landmark-to-image diffusion focuses on inpainting the lip region with explicit spatial cues from the previous predicted facial landmarks to create the image $\tilde{I}_i$ in the generated video. 
\ft{Given the normalized canonical landmarks $P_0^i$ predicted by the audio-to-landmark diffusion, we first denormalize them and restore their pose to that of the original corresponding frame.} Different from the previous audio-to-landmark diffusion, we draw the landmarks as an image $I_i^{P_0}$ to provide an explicit spatial structure relationship of the face expression.
By transforming the landmarks as images, the L2I diffusion process can render the target portrait image by querying the corresponding landmark spatially through self-attention interactions in unet. 

\noindent\textbf{Spatial-correspondence condition.}~\ft{Similar to DiffTalk~\cite{shen2023difftalk}, we also provide a random image in the same video as a reference image $I_{ref}$, which can provide the appearance cues for mouth region during generation. In the inference phrase, we will take the first frame as the reference image in all frames generation.}
{In addition to the reference image $I_{ref}$, we also provide its real landmark image $I_{ref}^P$ as a condition.} Given the reference image and the reference landmark image, the L2I diffusion model in the second hierarchy can effectively learn the spatial correspondence between the facial expression and the landmarks. Based on the learned spatial correspondence, the L2I diffusion model can inpaint the missing lip region by querying the target landmark image $I_i^{P_0}$, which is also a condition. Therefore, we utilize a target landmark $I_i^{P_0}$, a target masked image $I_i^{m}$, a reference image $I_{ref}$, and a reference landmark $I_{ref}^P$ to construct a spatial-correspondence condition for our landmark-to-image diffusion process.
\begin{figure*}[t]
  \centering
    \includegraphics[width=1\linewidth]{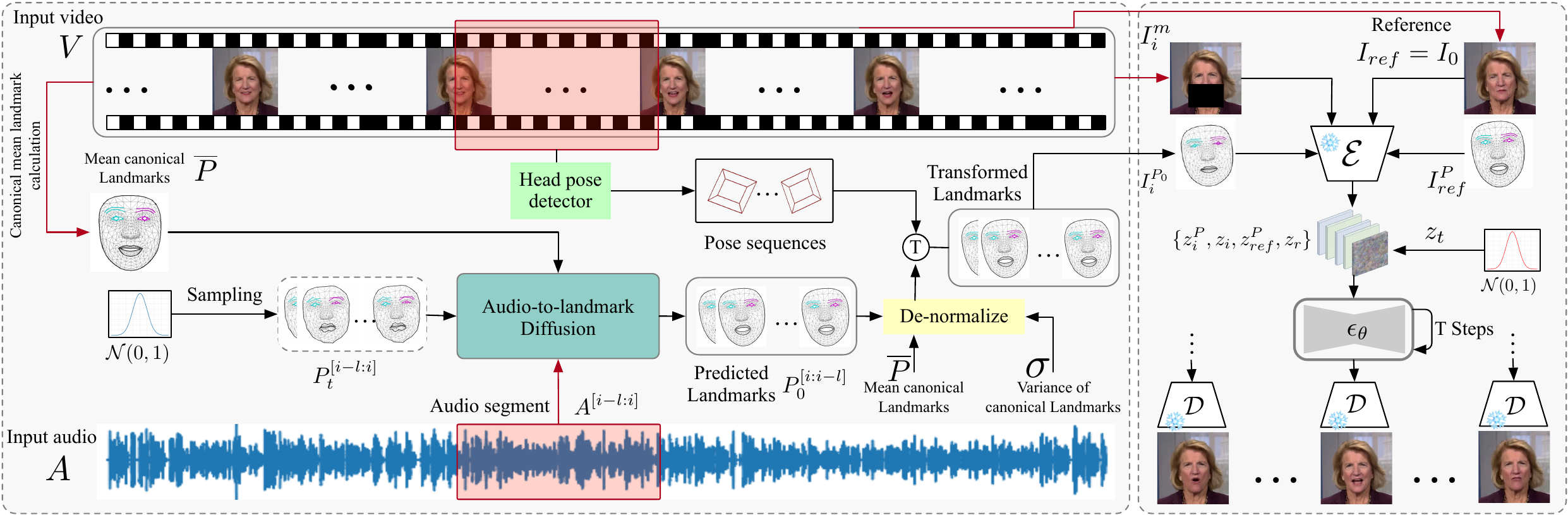}
    \vspace{-20pt}
    \caption{The illustration of the inference process. Given a video as a source and a segment of the input audio, {our DreamHead outputs a sequence of portrait images. ``$\textcircled{\scriptsize{T}}$'' means the transformation operation. We transform the the predicted landmarks $P_0^{[i,i-l]}$ after de-normalization. In this work, we use 3DFFA-v2~\cite{guo2020towards} as the head pose detector. $\sigma$ is the variance of the canonical landmarks (see Sec~\ref{sec:a2l} for details).
    }
    \vspace{-15pt}
}
    \label{fig:inference}       
\end{figure*}

\noindent\textbf{Diffusion learning.}~We adopt a latent diffusion model~\cite{rombach2022high} in our landmark-to-image diffusion process to produce portrait videos. 
{The image encoder $\mathcal{E}$ and the decoder $\mathcal{D}$ are pretrained in \cite{Esser_Rombach_Ommer_2021} and frozen during the training of our generation model.} Our conditions, including the masked target face image $I_i^{m} \in \mathbb{R}^{H\times W\times 3}$, the target landmark image $I_i^{P_0} \in \mathbb{R}^{H\times W\times 3}$, the reference image $I_{ref}\in \mathbb{R}^{H\times W\times 3}$, and the reference landmark image $I^P_{ref}\in \mathbb{R}^{H\times W\times 3}$, can be encoded into a latent space as:
\begin{equation}
    z_i^{m}, z^{P}_i, z_r, z_r^{P} := \mathcal{E}(I_i^{m}), \mathcal{E}(I^{P_0}_i),\mathcal{E}(I_{ref}),\mathcal{E}(I^P_{ref}),
\end{equation}
where $z_i^m$, $z_i^{P}$, $z_r$, and $z_r^{P}$ all have a dimension of $\mathbb{R}^{h \times w \times 3}$; $w$ and $h$ represent the spatial dimensions of latent codes; $H/h = W/w = f$; $f$ is a scale factor. In this way, the diffusion process can be conducted in a lower-dimensional latent space, which is more efficient requiring fewer computing resources. Given the above condition, we can model the learning of the landmark-to-image diffusion as follows:
\begin{equation}
    L_{l2i} := \mathbb{E}_{z,\epsilon\sim\mathcal{N}(0,1),t}\left[||\epsilon-L2I(z_t, t, C_{l2i})||\right],
\end{equation}
where L2I is our second hierarchy of landmark-to-image diffusion and $C_{l2i}$ is the condition set $\{z_i^m, z_i^{P},z_r,z_r^{P}\}$. All conditions are concatenated with the latent $z_t$ through the channel dimension and then fed into the second hierarchy of the landmark-to-image diffusion. After that, the spatial information from the the landmark image is explored by the intermediate self-attention layers of the landmark-to-image diffusion model. To this extent, we integrate all conditions into the diffusion network L2I to guide the talking head generation.

\vspace{-6pt}
\subsection{Training and Inference}
\vspace{-3pt}

\noindent\textbf{Training.} {The two  hierarchy diffusions in our DreamHead can be trained simultaneously to facilitate the optimization process}.
Generally, we utilize the pretrained Mediapiple model~\cite{lugaresi2019mediapipe} to extract dense facial landmarks from videos as shown in Figure~\ref{fig:audio2landmark}. We train our A2L diffusion and L2I diffusion simultaneously with landmarks as supervision and condition, respectively. Although the second hierarchy of landmark-to-image diffusion is trained on the ground-truth landmarks, our predicted landmarks from A2L diffusion is sufficient to provide an accurate lip shape. This is because the dense facial landmark representation inherently contains redundant information, which facilitates an effective approximation of the lip's actual distribution. 

\noindent\textbf{Inference.} \ft{As shown in Figure.~\ref{fig:inference}, given an audio and a video sequence as inputs, we capture and store the pose matrix for the face in each frame. Afterward, we transform the landmarks into the canonical space by detecting poses. We then calculate the mean of the canonical landmarks in all frames, denoted as $\overline{P}$, along with the variance of these landmarks. Subsequently, we employ the A2L diffusion process to predict temporal-consistency normalization of the canonical landmark sequences. Following this, we denormalize the predicted landmarks with the calculated mean and variance to restore the people's talking style. Then we transform the denormalized landmarks using the stored pose matrices to reconstruct the head motion. Lastly, we utilize the transformed landmarks as a condition to generate the final video, employing the L2I diffusion process on a frame-by-frame basis. It is notable that our method also accepts one image and audio as input. In this case, we will either randomly initialize a variance or use one from a random video and take the canonical landmarks of the given image as the mean landmarks $\overline{P}$.}


\begin{table*}[t]
\caption{Comparison with state-of-the-art methods. By using the landmark to learn the spatial-temporal correspondence, our method can produce portrait videos with accurate lip shapes and high Sync scores. }
\vspace{-10pt}
  \centering
  \resizebox{0.98\linewidth}{!}{
        \begin{tabular}{lccccc|ccccc}
        \toprule
        & \multicolumn{5}{c|}{HDTF dataset~\cite{zhang2021HDTF}} & \multicolumn{5}{c}{MEAD dataset~\cite{kaisiyuan2020mead}} \\
        \cline{2-6} \cline{7-11}
        
        Method & NIQE $\downarrow$ & LMD-$v$ $\downarrow$ & LMD $\downarrow$ & Sync $\uparrow$ & MA $\uparrow$ & NIQE $\downarrow$ & LMD-$v$ $\downarrow$ & LMD $\downarrow$ & Sync $\uparrow$ & MA $\uparrow$\\
        \midrule
        Ground Truth (reference) &6.38&0.00&0.00&6.07&1.00& 6.64 &0.00&0.00 &6.47&1.00\\
        \midrule
        MakeItTalk~\cite{zhou2020MakeItTalk} &8.18&1.91&2.23&3.9&0.53&8.11&2.17&2.28&2.11&0.60 \\
        Wav2Lip~\cite{prajwal2020lip} &7.83&2.08&1.98&\underline{5.78}&0.51& 7.40& 1.65& 1.86&\textbf{8.58}&0.61 \\
        Wav2LIp-GAN~\cite{prajwal2020lip} &7.77& 2.01& 1.98 & {5.78} &0.51 &\underline{7.09}&\underline{1.54}&\underline{1.37}&\underline{8.37}& \underline{0.71} \\
        Audio2Head~\cite{wang2021audio2head} &6.95&2.15&1.81&3.20 &0.55&8.41&1.96&2.21&6.24&0.59 \\
        PD-FGC \cite{wang2022pdfgc} & 8.10&2.29&4.15& \textbf{6.60}&3.34 & 9.62 & 2.16& 4.81& 6.25& 0.36\\
        SadTalker~\cite{zhang2022sadtalker} &7.07&2.43&2.37&3.96&0.51&8.86&2.16&3.24& 5.43& 0.55\\
        DiffTalk~\cite{shen2023difftalk} &\underline{6.76}&\underline{1.81}&\underline{1.82}&3.11&\underline{0.58}&8.41&1.74&2.47&5.86&0.54 \\
        \midrule
        DreamHead (Ours) &\textbf{6.53}&\textbf{1.67}&\textbf{1.35}&{4.63}&\textbf{0.62}&\textbf{7.00} &\textbf{1.41}&\textbf{0.99}&6.31&\textbf{0.80 }\\
        \bottomrule
        \end{tabular}
}
\label{tab:comp-sota}
\end{table*}
\begin{figure*}[t]
  \centering
    \includegraphics[width=1\linewidth]{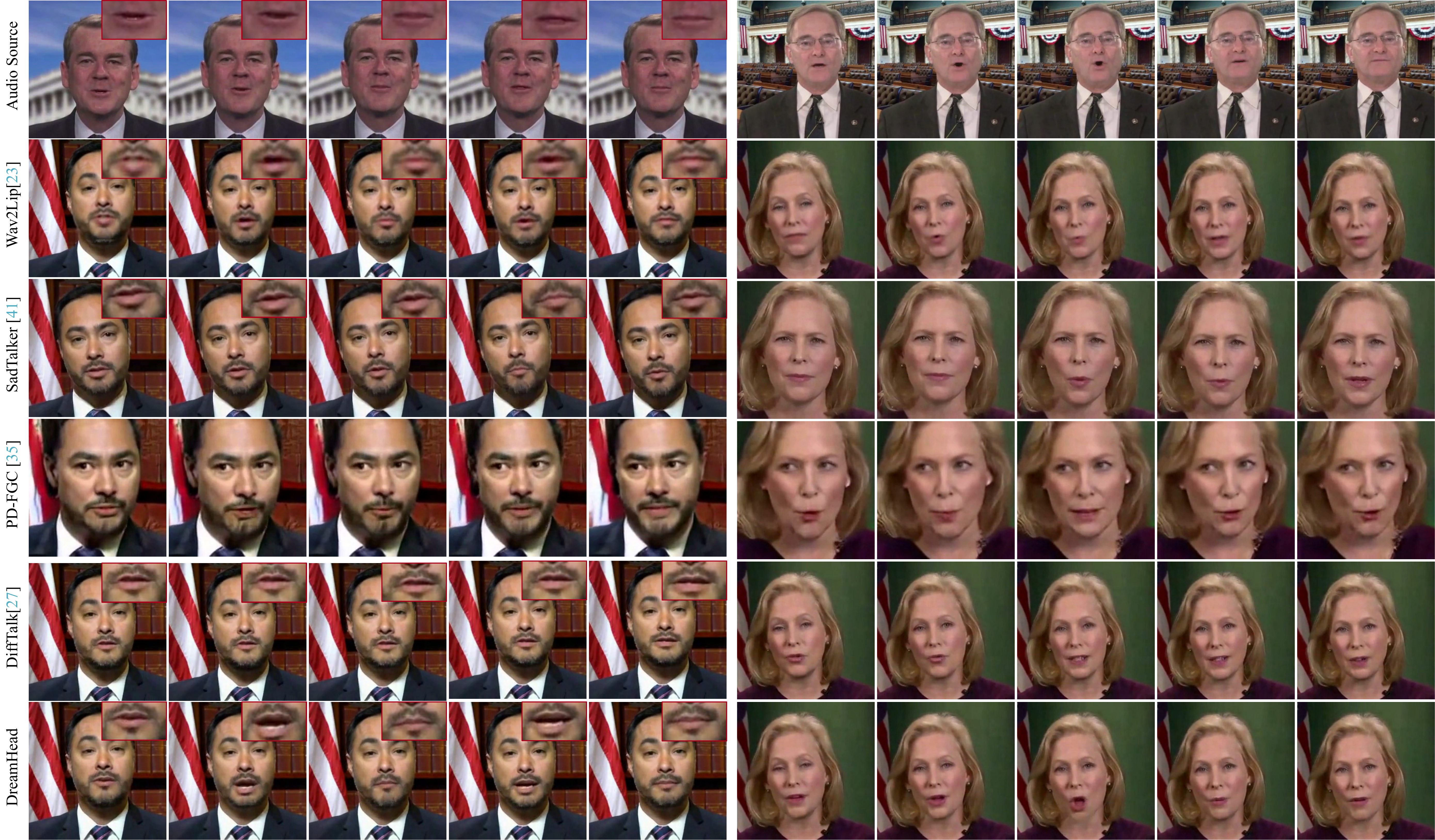}
    \caption{Visual comparison with other methods on cross-identity setting. Our method produces more accurate results compared with other methods.
    }
    \label{fig:comp-cross}       
\end{figure*}
\begin{figure*}[ht]
  \centering
    \includegraphics[width=1\textwidth]{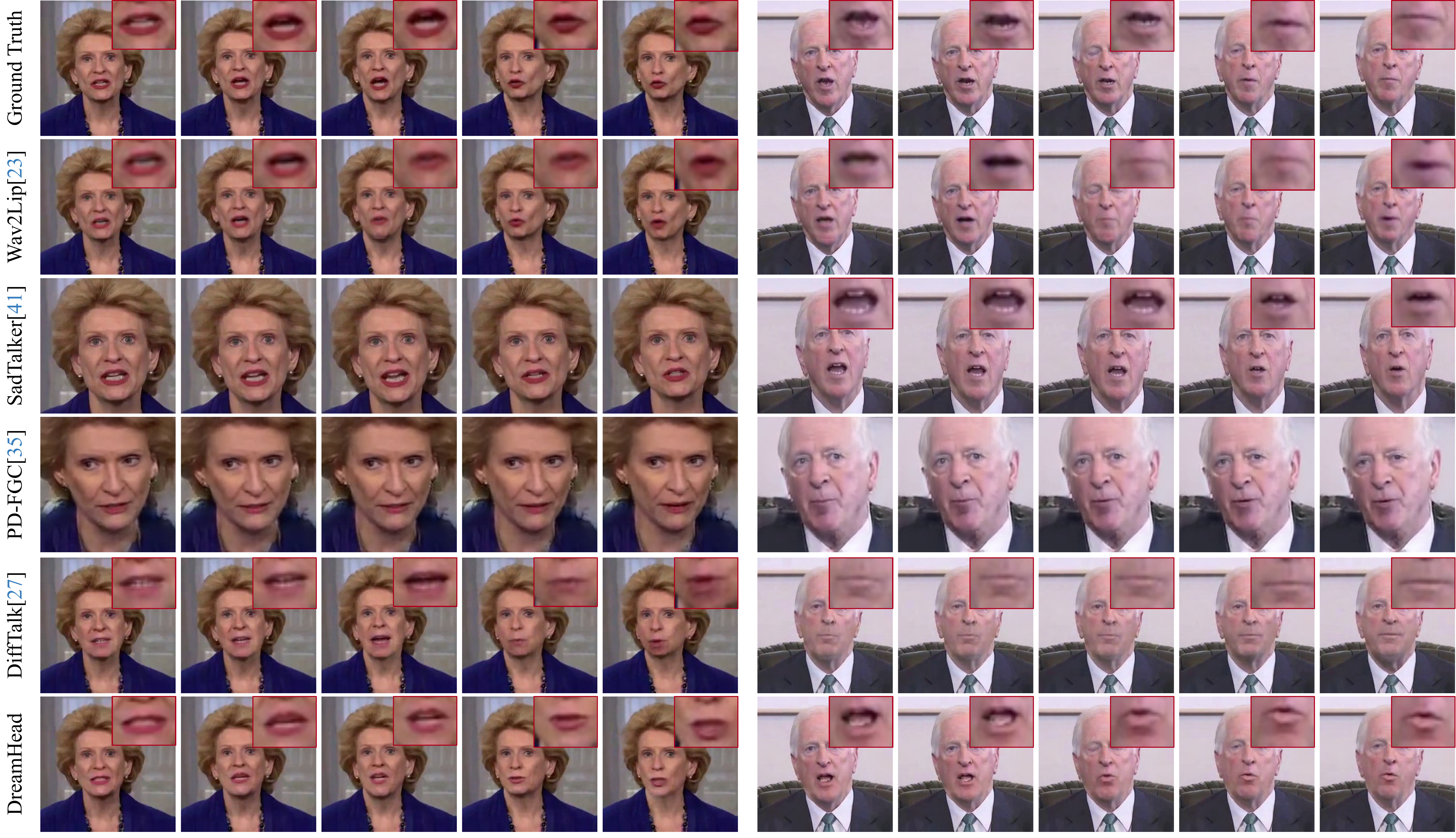}
    \caption{Visual comparison with other methods, our method produces more accurate and smooth results with high quality.
    }
    \vspace{-15pt}
    \label{fig:comp-sota-hdtf}       
\end{figure*}
\vspace{-6pt}

\section{Experiments}
\vspace{-3pt}
In this section, we present quantitative and qualitative experiments to evaluate our proposed DreamHead. More information and experimental results are reported in \emph{Supplementary Material.}
\vspace{-6pt}
\subsection{Experimental Settings}
\vspace{-3pt}
\noindent\textbf{Datasets.}
We evaluate our DreamHead on two audio-driven talking head generation datasets, \ie, HDTF~\cite{zhang2021HDTF} and MEAD~\cite{wang2020mead} dataset. For the HDTF dataset, we adopt the test set sampling strategy outlined in MODA~\cite{liu2023moda}. For the MEAD dataset, we randomly select 80\% videos as the training set and the remaining videos as the test set.

\noindent\textbf{Evaluation Metrics.}
Following MODA~\cite{liu2023moda}, 
we employ two key metrics for mouth accuracy assessment in our synthesized videos: Mouth Landmark Distance (LMD) and LMD velocity (LMD-$v$). We also compute Intersection-over-Union (IoU) to gauge overlap between predicted and ground truth mouth areas (MA). For audio-video synchronization, we rely on SyncNet's (Sync) confidence score~\cite{prajwal2020lip}. Additionally, we use the Natural Image Quality Evaluator (NIQE)~\cite{Mittal_Soundararajan_Bovik_2013} as our image quality metric. To assess landmark quality, we utilize ErrorNorm to evaluate positional accuracy and measure jitter between consecutive landmarks to assess their smoothness.

\vspace{-3pt}
\subsection{Implementation Details}
\vspace{-3pt}

In this work, we utilize Mediapipe\footnote{\url{https://google.github.io/mediapipe/}} to detect $478$ 3D facial landmarks for all videos. We resize the input image in the landmark-to-image diffusion process to $256 \times 256$, and set the downsampling factor $f$ at $4$. The length of the denoising step, $T$, is set to $1000$ for both the audio-to-landmark and landmark-to-image diffusion processes. In the first hierarchy of the audio-to-landmark diffusion, the number of temporal blocks is set to $12$. For the audio-to-landmark diffusion, we set $l$ to $300$ and $20$ for the HDTF and MEAD datasets, respectively, to account for the longer video lengths in the HDTF dataset. We set the frame interval $\tau$ in the landmark-to-image diffusion to $20$ to prevent the model from learning a shortcut with frames that are too close together.
\vspace{-6pt}
\subsection{Comparison with state-of-the-art methods}
\vspace{-3pt}

In Table~\ref{tab:comp-sota}, we compare our methods with other SOTA subject-general methods. As shown in Table~\ref{tab:comp-sota}, our DreadHead can achieve the best performance on all datasets. 
{Because of designing the diffusion model for effective learning of spatial-temporal correspondences, our method can produce the highest quality video with $6.53$ NIQE on the HDTF dataset.}
Regarding the correctness of the audio-lip synchronization, our method also shows the best performance. \ie, DreamHead obtains the lowest LMD and LMD-$v$, and the highest MA scores on the two datasets. {Because our portrait video is rendered by the condition of dense facial landmarks, DreamHead can synthesize accurate lip shapes}. Since Wav2Lip~\cite{prajwal2020lip} takes the score from the SyncNet~\cite{chung2017lip} as supervision during their training, Wav2Lip and Wav2Lip-GAN present the highest Sync scores which are even higher than the ground-truth in MEAD dataset~\cite{wang2020mead}. \ft{Our lower sync score is due to our adherence to the unique talking styles of test subjects, such as those who speak with almost closed mouths, which naturally results in lower sync scores even in real videos}. However, our method still obtains a better sync score compared with most of compared methods. These results demonstrate that accurate facial landmarks can benefit the generation of lip-synced portrait videos using diffusion models.
Additionally, we also visualize some samples to demonstrate the performance of our DreamHead. As shown in Figure~\ref{fig:comp-sota-hdtf} and Figure~\ref{fig:comp-cross}, we can observe that our method can produce a more accurate and smooth portrait video, because we model the spatial correspondence between facial landmarks and real face images. Compared with Wav2Lip~\cite{prajwal2020lip}, our result maintains high fidelity because of the proposed hierarchical diffusion. These results verify that our method can effectively learn the spatial-temporal correspondence to boost the generation performance with the diffusion model. 

We conduct user studies with 20 attendees on 40 videos generated by ours and the other methods. Each participant is asked to select the best generated talking-portrait videos based on two major aspects: lip synchronization accuracy and quality of the generated video.
The statistics are reported in Table~\ref{tab:user_study}. Overall, users prefer our results on lip synchronization accuracy and quality. The results indicate the effectiveness of the proposed method.

\begin{table}[t]
\begin{minipage}[tb]{0.48\textwidth}
\centering
\caption{User study. Users prefer results generated by our method}
\vspace{-10pt}
  \resizebox{1\linewidth}{!}{
        \begin{tabular}{lcccc}
        \toprule
        Method & Wav2Lip & SadTalker & DiffTalk & DreamHead(Ours)   \\
        \midrule
        Correctness & 28.57\%& 5.76\% &9.52\% & \textbf{57.14\%} \\
        Quality & 19.05 \%& 14.28 \%& 11.90\% & \textbf{54.76\%} \\
        \bottomrule
        \end{tabular}
}
\label{tab:user_study}
\end{minipage}
\begin{minipage}[tb]{0.48\textwidth}
\centering
\caption{Temporal-consistency comparison of talking head diffusion model.}
\vspace{-10pt}
  \resizebox{1\linewidth}{!}
  {
        \begin{tabular}{lcc}
        \toprule
        Method & TCM $\uparrow$ (HDTF~\cite{zhang2021HDTF}) & TCM $\uparrow$ (MEAD~\cite{wang2020mead})\\
        \midrule
        DiffTalk~\cite{shen2023difftalk} & 0.163& 0.173\\
        DreamHead (Ours) &\textbf{0.179}&\textbf{0.186}\\
        \bottomrule
        \end{tabular}
}
\label{tab:comp-tcm}
\end{minipage}
\vspace{-10pt}
\end{table}

\begin{figure*}[t]
  \centering
    \includegraphics[width=1\textwidth]{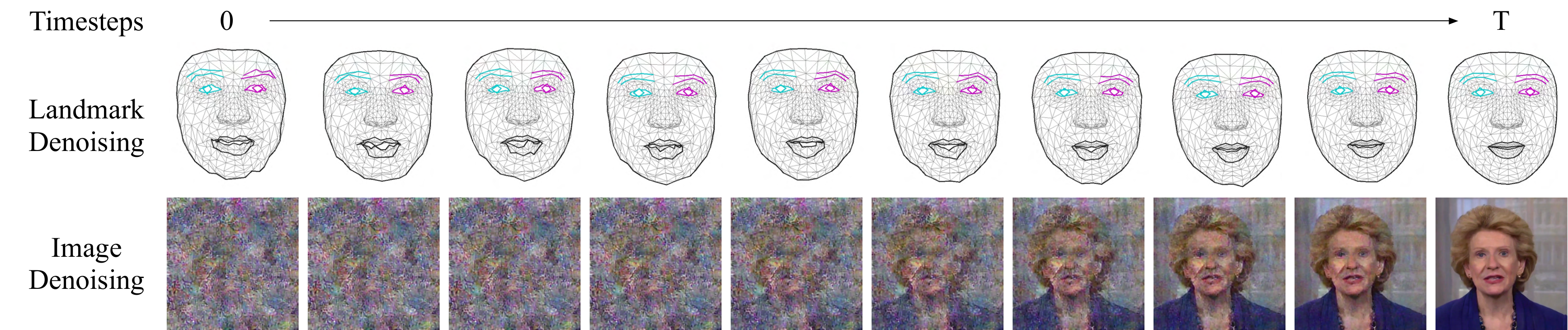}
    \vspace{-20pt}
    \caption{Denoising process of DreamHead. Using diffusion processes, we can produce smooth and accurate landmarks to provide spatial information for image generation.
    }
    \vspace{-15pt}
    
    \label{fig:a2l-denoise}       
\end{figure*}

\begin{figure}[t]
  \centering
    \includegraphics[width=1\linewidth]{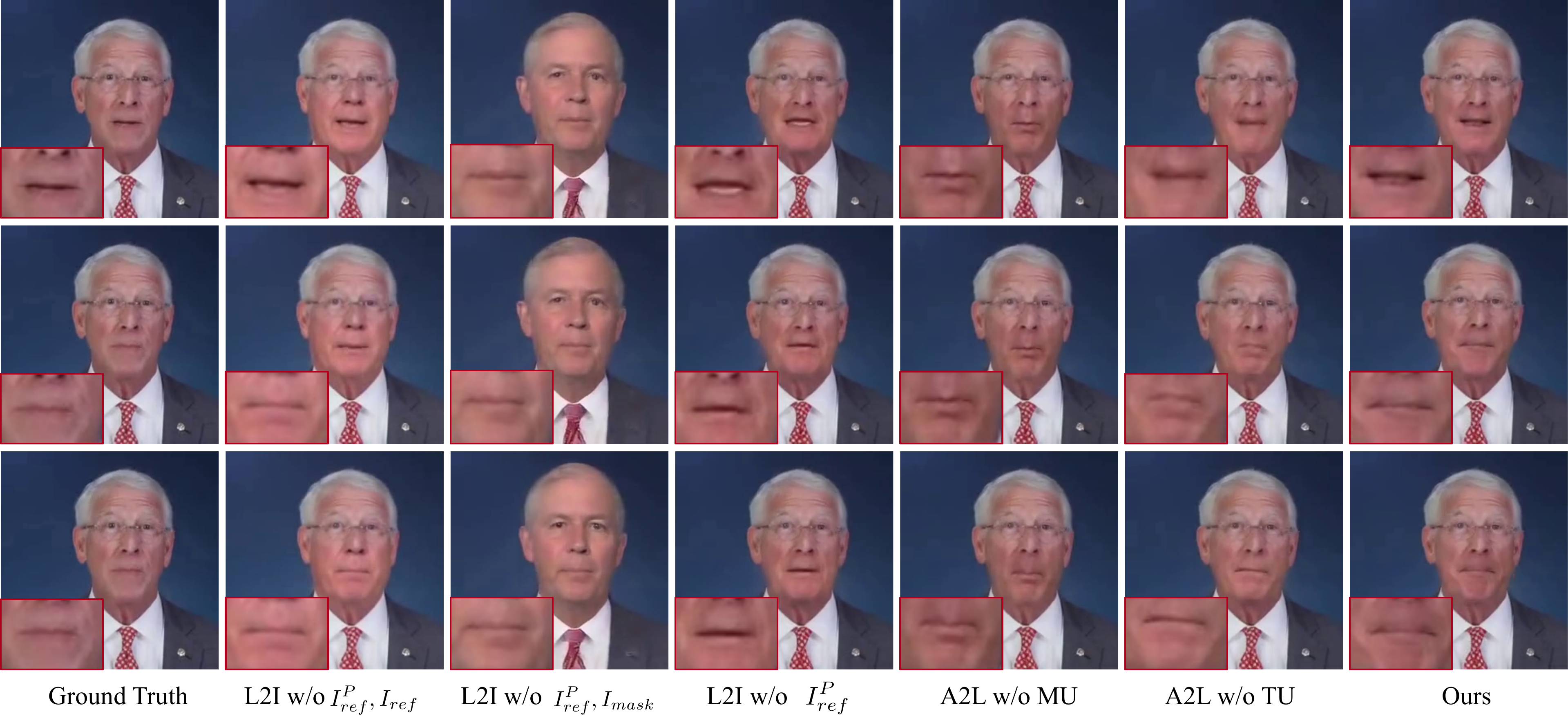}
    \vspace{-20pt}
    \caption{Ablation study on different components. We select three consecutive frames to check the accuracy of lip-syncing and the temporal consistency.
    }
    \vspace{-15pt}
    \label{fig:abla}       
\end{figure}
\vspace{-6pt}
\subsection{Ablation studies}
\vspace{-3pt}

\noindent\textbf{Visualization of estimated landmarks.} Firstly, we demonstrate that our method can predict dense facial landmarks given the audio cues. As shown in Figure~\ref{fig:firstpage}, our first hierarchy of audio-to-landmark diffusion can produce lip-synced and jittering-less landmark sequences, which can provide explicit spatial information of the lips during image generation. Furthermore, we also visualize the denoise process of the landmarks in Figure~\ref{fig:a2l-denoise}. By employing the landmark normalization technique, the initial landmark ($t=0$) still maintains a rough facial shape after de-normalization. Finally, our audio-to-landmark diffusion produces a reliable landmark image after multiple denoising steps. It demonstrates the effectiveness of using our diffusion model to estimate facial landmarks.

\noindent\textbf{Temporal consistency in the diffusion model.} Temporal consistency is a key challenge in the diffusion model. In this work, we compare the temporal consistency with other sota diffusion-based generic talking head methods, \eg DiffTalk~\cite{shen2023difftalk}. We utilize the temporal consistency metric to measure the frame-wise consistency (TCM~\cite{2020Unsupervised})\footnote{The detail of the TCM is reported in the \emph{Supplementary Metarial}} of the generated portrait video. As shown in Table~\ref{tab:comp-tcm}, our method can produce more temporal-smooth results on two datasets compared with DiffTalk~\cite{shen2023difftalk}. These results validate that using the landmark as an intermediate representation in our diffusion framework can constrain the temporal consistency of the generated video.


\noindent\textbf{Diffusion for landmark estimation.} Compared with the traditional CNN network, our diffusion framework can produce dense facial landmarks with fewer jittering artifacts. To verify this, we use the audio-to-landmark network to directly predict the landmarks given the input of mean facial landmark and audio sequence (``A2L w/o diffusion'' in Table~\ref{tab:abla-ldm}). From Table~\ref{tab:abla-ldm}, we can observe that our audio-to-landmark network trained in a diffusion manner can produce fewer jittering artifacts compared with ``A2L w/o diffusion''. 
Notably, our method smooths out some of the outlier landmarks within the sequence, which results in an increased ErrorNorm. These outlier landmarks can potentially align with the ground truth due to potential errors in predicting the ground truth with a pre-trained model. Furthermore, we evaluate the effectiveness of the temporal unit (TU) and mapping unit (MU) in the audio-to-landmark diffusion. The results shown in Table~\ref{tab:abla-ldm} demonstrate that the temporal unit plays a critical role in maintaining the temporal consistency (\ie, 5.848 vs. 2.646 for Jitter on HDTF dataset). By integrating the temporal unit and the mapping unit, our first hierarchy of audio-to-landmark diffusion can estimate accurate landmarks with fewer jittering artifacts.

\begin{table}[t]
\begin{minipage}[tb]{0.48\textwidth}
\centering
\caption{\scriptsize{Ablation study on the quality of estimated landmark. By using the diffusion process to align the landmark with audio gradually, the estimated landmarks become more smooth.}} 
\vspace{-10pt}
  \centering
  \resizebox{1\linewidth}{!}{
  \renewcommand{\arraystretch}{1.38}
        \begin{tabular}{lcccc}
        \toprule
        & \multicolumn{2}{c}{HDTF dataset~\cite{zhang2021HDTF}} & \multicolumn{2}{c}{MEAD dataset~\cite{kaisiyuan2020mead}} \\
        \cline{2-5}
        Method & ErrorNorm $\downarrow$ & Jitter $\downarrow$ & ErrorNorm $\downarrow$ & Jitter $\downarrow$  \\
        \midrule
        A2L w/o TU & 3.020 & 5.848 & 2.836& 3.394\\
        A2L w/o MU &2.938 & \underline{2.695}& 2.972 & \underline{1.685}\\
        A2L w/o diffusion & \textbf{2.189}& 4.490& \textbf{2.238} &2.470 \\
        \midrule
        DreamHead (Ours) & \underline{2.569}&\textbf{2.646} & \underline{2.403}&\textbf{1.641}\\
        \bottomrule
        \end{tabular}
}
\label{tab:abla-ldm}
\end{minipage}
\begin{minipage}[tb]{0.48\textwidth}
\centering
\caption{\scriptsize{Ablation study. These results verify that estimating smooth and accurate landmarks to learn the spatial-temporal correspondence can benefit image generation.}}
\vspace{-10pt}
  \centering
  \resizebox{1\linewidth}{!}{
        \begin{tabular}{lccccc}
        \toprule
     
        Method & NIQE $\downarrow$ & LMD-$v$ $\downarrow$ & LMD $\downarrow$ & Sync $\uparrow$ & MA $\uparrow$ \\
        \midrule
        L2I w/o $I^P_{ref}$,$I_{ref}$ & 6.57& 1.67& 1.39& 4.40& 0.62\\
        L2I w/o $I^P_{ref}$,$I_i^{m}$ & 7.10& 1.72& 1.39&4.39&0.62\\
        L2I w/o $I^P_{ref}$ & 6.53&1.69&1.36&4.44&0.63\\
        \midrule
        A2L w/o TU & 6.52 & 2.27& 1.53 & 2.40 & 0.60 \\
        A2L w/o MU & 6.54& 1.68 & 1.44 & 3.27 & 0.61 \\
        \midrule
        Ground Truth (reference) &6.38&0.00&0.00&6.07&1.00\\
        DreamHead (Ours) &\textbf{6.53}&\textbf{1.67}&\textbf{1.35}&\textbf{4.63}&{0.62}\\
        \bottomrule
        \end{tabular}
}
\label{tab:abla}
\end{minipage}
\vspace{-20pt}
\end{table}

\noindent\textbf{The components in A2L diffusion.} We also evaluate the influence of the temporal unit and the mapping unit on the final generated portrait video. As shown in Table~\ref{tab:abla}, the audio-to-landmark model produces poor results if without the temporal unit and mapping unit. 
{We can observe that TU and MU mainly affect lip synchronization. The quality of the generated images of ``A2L w/o TU'' and ``A2L w/o MU'' is comparable to our full method (see NIQE values in Table~\ref{tab:abla}), because our L2I diffusion can ensure high-quality results using landmarks.}
We also visualize some samples to verify the effectiveness of each component in A2L diffusion in Figure~\ref{fig:abla}. As shown in Figure~\ref{fig:abla}, the model with both the temporal and mapping units can achieve pose-smooth video results. Without the temporal unit, the mouth region changes dynamically. It verifies that our designed temporal and mapping units indeed contribute to the temporal consistency of our final generated videos.

\noindent\textbf{Spatial-correspondence condition in L2I diffusion.} In this paper, we propose a hierarchical diffusion to learn the spatial-temporal correspondence to produce spatially consistent and temporally smooth portrait videos. We construct a spatial-correspondence conditions in L2I diffusion.
There are four conditions in our L2I diffusion, including the target masked image $I_i^{m}$, the target landmark $I_{i}^P$, the reference landmark $I_{ref}^P$, and the reference image $I_{ref}$. From Table~\ref{tab:abla}, we can observe that adding the reference landmark as a condition can boost the performance via a comparison between ``L2I w/o $I^P_{ref}$'' and full method ($4.44$ vs. $4.63$ on Sync score). With both $I_{ref}^P$ and $I_{ref}$, the diffusion model is more effective to learn the spatial correspondence between the facial appearance and the landmarks. Furthermore, the target masked image provides identity information for the generation. As shown in Figure~\ref{fig:abla}, without $I_i^{m}$, the identity of generated results is different from the ground truth (``L2I w/o $I_{ref}^P, I_i^{m}$'' vs. ``L2I w/o $I_{ref}^P$''). In Figure~\ref{fig:abla}, our full method generates more accurate and smooth results. Moreover, we also show the denoise process of images in Figure~\ref{fig:a2l-denoise}. We can gradually generate images with high fidelity by the diffusion process. {These results in Table~\ref{tab:abla} and Figure~\ref{fig:abla} demonstrate that our DreamHead can produce lip-synced images based on the learned spatial-temporal correspondences.}

\vspace{-6pt}

\section{Conclusion}
In this paper, we proposed a hierarchical diffusion framework, DreamHead, to learn spatial-temporal correspondence using diffused landmarks as intermediate representations. The first hierarchy of audio-to-landmark diffusion leverages the correlation between audio cues and facial landmarks to produce temporal-smooth landmark sequences. The second hierarchy of landmark-to-image diffusion learns the spatial correspondence between landmarks and face expressions.
Extensive results clearly demonstrate that our DreamHead can estimate effective spatial-temporal correspondences between the input audio and the output face image, creating new state-of-the-art performances.
{
    \small
    \bibliographystyle{ieeenat_fullname}
    \bibliography{main}
}


\end{document}


\twocolumn[{%
\renewcommand\twocolumn[1][]{#1}%
\maketitle
\begin{center}
    \centering
    \captionsetup{type=figure}
    \vspace{-8pt}
    \includegraphics[width=0.99\textwidth]{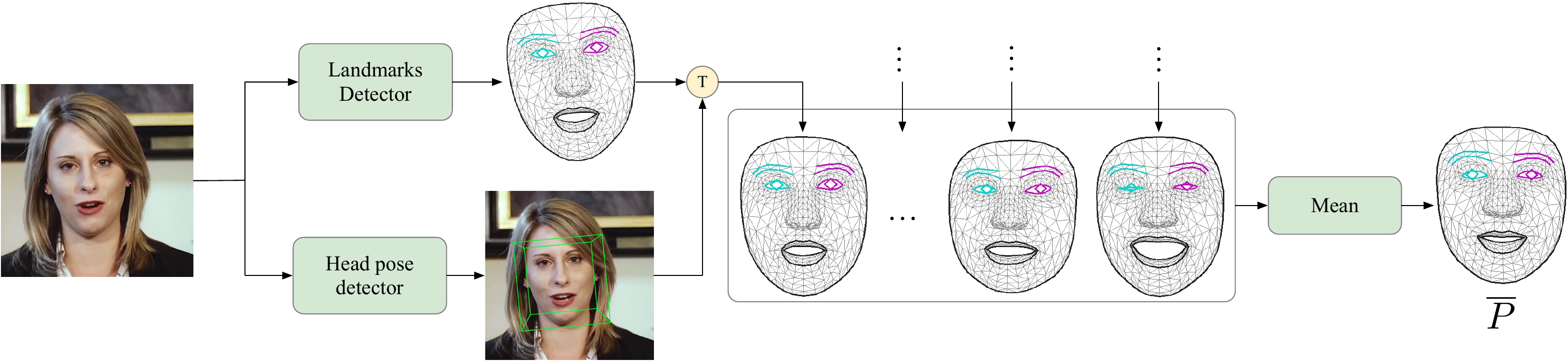}
    \vspace{-5pt}
    \captionof{figure}{The illustration of data pre-processing. We utilise a pre-trained  Mediapipe~\cite{lugaresi2019mediapipe} model to detect the dense facial landmarks for each face and a 3DFFA-v2~\cite{guo2020towards} to detect the head pose. ``T'' represents the transformation operation.
    }
    \label{fig:preprocess}
\end{center}%
}]

\section{Data Pre-process}
In this work, we further preprocess the dense facial landmarks detected by the Mediapipe model~\cite{lugaresi2019mediapipe} to enhance the connections between audio cues and facial landmarks by filtering out audio-irrelevant information from the dense facial landmarks. As shown in Figure~\ref{fig:preprocess}, we first utilize the Mediapipe model~\cite{lugaresi2019mediapipe} to detect dense 3D facial landmarks for each face image. To transform these 3D landmarks to a canonical space, we apply a pose detector (e.g., 3DFFA-v2~\cite{guo2020towards}) to obtain the head pose of the face. Subsequently, we obtain a sequence of 3D landmarks for each video. Then, we use all canonical landmarks from a video to calculate the mean landmarks, denoted as $\overline{P}$, for that video.
\section{Implementation Details}
In this work, we utilize Mediapipe\footnote{\url{https://google.github.io/mediapipe/}} to detect $478$ 3D facial landmarks for all videos. We resize the input image in the landmark-to-image diffusion process to $256 \times 256$, and set the downsampling factor $f$ at $4$. The length of the denoising step, $T$, is set to $1000$ for both the audio-to-landmark and landmark-to-image diffusion processes. In the first hierarchy of the audio-to-landmark diffusion, the number of temporal blocks is set to $12$. For the audio-to-landmark diffusion, we set $l$ to $300$ and $20$ for the HDTF and MEAD datasets, respectively, to account for the longer video lengths in the HDTF dataset. We set the frame interval $\tau$ in the landmark-to-image diffusion to $20$ to prevent the model from learning a shortcut with frames that are too close together.

\begin{figure*}[t]
  \centering
    \includegraphics[width=1\linewidth]{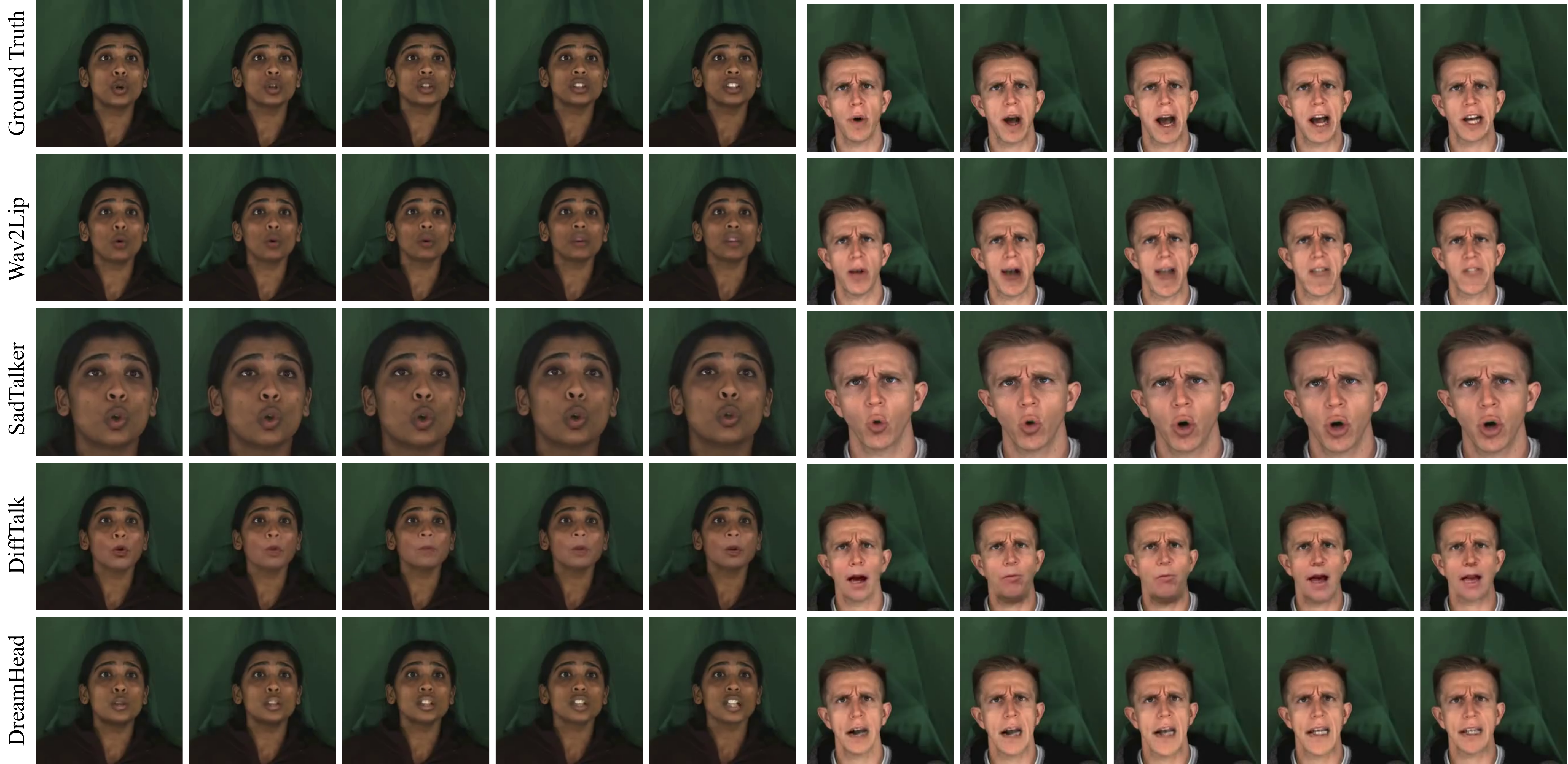}
    \caption{Visual comparison with other methods on the MEAD dataset~\cite{kaisiyuan2020mead}. We selected five consecutive frames for each sample. Our method produces more accurate results compared to other methods.
    }
    \vspace{-5pt}
    \label{fig:comp-sota-mead}       
\end{figure*}

\section{More Experiments}
\begin{table}[t]
\caption{Evaluation on the frames interval.}
  \centering
  \resizebox{1\linewidth}{!}{
        \begin{tabular}{lccccc}
        \toprule
     
        Method & NIQE $\downarrow$ & LMD-$v$ $\downarrow$ & LMD $\downarrow$ & Sync $\uparrow$ & MA $\uparrow$ \\
        \midrule
        Interval = 10 & 6.58& 1.68&1.35&4.57&0.62 \\
        Interval = 15 & 6.55&1.65&1.35&4.78&0.63\\
        Interval = 25 & 6.54&1.66&1.37&4.58&0.62 \\
        Interval = 30 &6.54&1.65&1.35&4.75&0.63\\
        Interval = 40 &6.55&1.65&1.36&4.73&0.62\\
        Random &6.55&1.70&1.37&4.63&0.62\\
        
        \midrule
        Ground Truth (reference) &6.38&0.00&0.00&6.07&1.00\\
        Interval=20 (Ours) &\textbf{6.51}&\textbf{1.65}&\textbf{1.35}&\textbf{4.90}&\textbf{0.63}\\
        \bottomrule
        \end{tabular}
}
\label{tab:abla-interval}
\end{table}
\subsection{Hyper-parameters Evaluation}
\noindent\textbf{Frames interval.} In the spatial-temporal condition within the second hierarchy of the landmark-to-image diffusion process, we select the $(i-\tau)$-th frames in the video as a reference image when rendering the $i$-th frames. This approach provides more temporal information for lip synchronization. We evaluate the effectiveness of the interval $\tau$ in our work. The results are shown in Table~\ref{tab:abla-interval}. According to Table~\ref{tab:abla-interval}, cases with fixed intervals produce comparable results, while randomly selecting reference frames can lead to a mismatch in lip velocity due to missing temporal information. We have chosen $\tau=20$ as it achieves the best comprehensive results.

\begin{table}[h]
\caption{The evaluation of the number of temporal blocks in our audio-to-landmark diffusion. }
  \centering
  \resizebox{1\linewidth}{!}{
        \begin{tabular}{lccccc|c}
        \toprule
     
        Method & N=3 & N=6 & N=9& N=15 & N= 18& Ours(N=12) \\
        \midrule
        ErrorNorm $\downarrow$ &2.994&2.734&2.783&2.599&2.623&2.569 \\
        Jitter $\downarrow$ &2.652&2.653&2.671&2.648&2.647&2.646\\
        \bottomrule
        \end{tabular}
}
\label{tab:abla-tb-num}
\end{table}
\noindent\textbf{The number of temporal blocks.} In our study, we assessed the impact of varying the number of temporal blocks in our audio-to-landmark diffusion model, with the findings detailed in Table~\ref{tab:abla-tb-num}. An analysis of the data in Table~\ref{tab:abla-tb-num} reveals that the model's performance improves as the number of temporal blocks increases. However, this improvement plateaus when the number of blocks reaches or exceeds $12$ ($N \geq 12$). Consequently, to optimize computational efficiency without compromising performance, we implemented the model with $N = 12$ temporal blocks.

\subsection{More Visualization}
\noindent\textbf{Comparision on MEAD dataset.} As shown in Figure~\ref{fig:comp-sota-mead}, we selected five consecutive frames from each sample to assess temporal consistency and lip synchronization. According to Figure~\ref{fig:comp-sota-mead}, Wav2Lip~\cite{prajwal2020lip} produces low-quality results, as they are trained at a low resolution to ensure the training stability of their lip expert. Meanwhile, SadTalker~\cite{zhang2022sadtalker} and DiffTalk~\cite{shen2023difftalk} yield inaccurate results. In contrast, our DreamHead method produces the best results, as it learns spatial information from diffused facial landmarks, which are temporally aligned with audio cues.
{
    \small
    \bibliographystyle{ieeenat_fullname}
    \bibliography{main}
}
